\RequirePackage{fix-cm}
\documentclass[natbib,smallextended]{svjour3}       
\smartqed  
\usepackage{hyperref}

\hypersetup{colorlinks=true,citecolor=blue,urlcolor=blue}
\journalname{Computational Economics}
\usepackage{enumerate}
\usepackage{algorithm}
\usepackage{algpseudocode}
\usepackage{algorithmicx}
\usepackage{fancyhdr}
\usepackage{amsmath}
\usepackage[pdftex]{graphicx}
\usepackage{epstopdf}
\begin{document}
\title{A New Methodology for Estimating Internal Credit Risk and Bankruptcy Prediction under Basel II Regime}
\titlerunning{A New Methodology for Estimating Credit Risk and Bankruptcy Prediction}
\author{M. Naresh Kumar$^1$ \thanks{ $^1$ Corresponding author: M. Naresh Kumar, National Remote Sensing Centre (ISRO), Hyderabad - $500 037$, A.P., INDIA. Tel.: +91 40 23884388 \email{nareshkumar\_m@nrsc.gov.in }}  \and V. Sree Hari Rao$^2$  \thanks{ $^2$ V. Sree Hari Rao, Institute for Development and Research in Banking Technology, Masab Tank, Hyderabad, Andhra Pradesh, $500 057$, India. Present Address : Foundation for Scientific Research and Technological Innovation (FSRTI), Alakapuri, Hyderabad- $500 035$, A.P., INDIA. Tel.: +91 40 24038943 \email{vshrao@gmail.com}}
\thanks{ $^2$ The work was supported by the Foundation for Scientific Research and Technological Innovation (FSRTI)- A Constituent Division of Sri Vadrevu Seshagiri Rao Memorial Charitable Trust, Hyderabad - 500 035, India under grant \# FSRTI/R.P-1/2012-13.}}
\institute{}					

\date{Received: 30-Apr-2013/Accepted: 07-July-2014/Published:27-July-2014} 
\maketitle
\markboth{\scriptsize $\copyright$ Springer US. Published in Computational Economics. DOI: 10.1007/s10614-014-9452-9} {\scriptsize \thepage}
\begin{abstract}
Credit estimation and bankruptcy prediction methods have utilized Altman's Z-score method for the last several years. It is reported in many studies that Z-score is sensitive to changes in accounting figures. Researchers have proposed different variations to conventional Z-score that can improve the prediction accuracy. In this paper, we develop a new multivariate nonlinear model for computing the Z-score. In addition, we develop a new credit risk index by fitting a Pearson type 3 distribution to the transformed financial ratios. The results of our study have shown that the new Z-score can predict the bankruptcy with an accuracy of $98.6\%$ as compared to $93.5\%$ by Altman's Z-score. Also, the discriminate analysis revealed that the new transformed financial ratios could predict the bankruptcy probability with an accuracy of $93.0\%$ as compared to $87.4\%$ using the weights of Altman's Z-score.
\end{abstract}

\section{Introduction}
Credit ratings have become an integral part of today's capital markets as they help in the evaluation and assessment of credit risk, benchmark issues and create secondary markets for those aspects. Credit risk exists virtually in all income-producing activities and their inappropriate evaluation or inadequate mitigation would result in failure of institutions. In general, the credit ratings given by agencies such as Standard and Poor, Moody and Fitch are based on the probability of default and recovery rate taking into account not only the variables in the financial statement of the firms but also the market cues. Predicting bankruptcy for firms in financial distress from the financial statement history is an important problem studied widely by the researchers \citep{37a,40a,45a,6a}. Among them the Altman's Z-score \citep{2,3,4,5} is the most popular and widely accepted metric for predicting the bankruptcy. The popularity of Z-score may be attributed to its simplicity in computation and ease in its application \citep{7,8,9,10}.

Altman's Z-score uses mainly accounting figures in the financial statement as variables in the computation. The Z-score is highly sensitive to small variations in these figures due to its dependency on them. This leads to an exaggerated Z-score in case they are manipulated as it does not include the past accounting profile of the business into consideration during its computation.  Therefore, the bankruptcy probability predictions using Altman's Z-score would cause significant levels of type-I errors (classifying bankrupt firms as non-bankrupt). \cite{11} demonstrated the bias of Z-score in predicting the bankruptcy. In addition, the models have the weakness of not being immune to false accounting practices \citep{17}. It is stated by Altman that the retained earnings account is subject to manipulation via corporate quasi-reorganizations and stock dividend declarations, which may cause a bias \citep{16}. Moreover the weights used by Altman are still prevalent even though the financial reporting environment has changed drastically from a rule-based approach to a principle-based set of standards aiming for harmonization with the International Financial Reporting Standards (IFRSs) \citep{12,13,14}.

The Z-score, which is a derivative of financial ratios may not represent different risks using same quantitative figures or same business risks for different financial statement figures. To account for this asymmetry, the Z-score may be adjusted  by including the earnings management in the computation procedure \citep{1} but it may also be exposed to manipulations in the accounting data. These models should not be applied to financial firms due to their frequent use of off balance-sheet items \citep{15}. Also, the results of the model may vary over time, which may be explained by the uncertainty of the stock prices as they are subject to the stock market opinion. During periods when the stock market is relatively high, the Z-score outcomes will be higher than in times when stock prices are low.

The financial scores are linearly combined to obtain the Z-score using the weight functions derived from the multivariate discriminate analysis. In realistic scenarios the financial ratios used as independent variables may not be linearly related. Also, the score is biased to small variations in the financial scores. Moreover, it is not possible to compare the performance of different firms such as non-manufacturing, manufacturing as the weights of the financial ratios would differ among different firms. Further, it is difficult/not feasible to develop specific models tailored to address the scenarios of each type of industry (retailers, airlines etc.) even though it may look ideal \citep{15}.

Keeping in view the shortcomings of Altman's Z-score and the adjusted Z-score \citep{1}, we frame the following objectives for the present study:
\begin{enumerate}

\item developing a score based method using a nonlinear form of financial ratios;
 \item designing an index using an equi-probability transformation by fitting a Pearson type 3 (P3) distribution to the newly developed Z-score say $Z_M$;
 \item formulating a rating scheme based on the index;
 \item comparing the $Z_M$ with Altman's Z-score and the proposed rating scheme with those being employed widely by financial institutions for bankruptcy predictions.

\end{enumerate}

The present paper is organized as follows. In Section~\ref{sec2} we propose a new nonlinear transformation model for computing the $Z$-score while in Section~\ref{sec3} a new index (based on the $Z_M$) using a P3 distribution is developed. The methodology followed for predicting the bankruptcy probability of firms is presented in Section~\ref{sec4}. The datasets and the results are described in Section~\ref{sec5}. Conclusions and discussion are deferred to Section~\ref{sec6}.

\section{A generalized nonlinear score for modeling the financial ratios}
\label{sec2}
The standard Z-score is a statistical measurement of a scores's relationship to the mean in a group of scores generally measured by the formula
\begin{equation}
Z=\frac{(X-\mu)}{\sigma},
\end{equation}
in which $X$ denotes the set of measurements, $\mu$ and $\sigma$ respectively denote the mean and standard deviation of the data in the set $X$. The Z-score is a very useful statistic for obtaining the probability of a score occurring within a normal distribution and comparison of two scores that are from different normal distributions. \cite{4} first proposed a Z-score measuring a company's financial strength using a weighted sum of several factors among the variables (financial ratios) that gives an approximate description of the bankruptcy probability.

\cite{4} utilized a data set composed of sixty-six corporations with thirty-three firms in each of the two risk groups and financial ratios given in Table~\ref{tab1} to obtain a set of ratios that influence the bankruptcy prediction. The mean asset size of these firms was \$6.4 million, with a range between \$0.7 million and \$25.9 million.
\begin{table}[ht]
\caption{Financial Ratios previously used in bankruptcy prediction studies}
\centering
\begin{tabular}{c l l}
\hline\hline
Ratio\#& Name & Type   \\ [0.5ex]
\hline
R1 & Cash/Current Liabilities& liquidity \\
R2 &Cash Flow/Current Liabilities& liquidity\\
R3 &Cash Flow/Total Assets &liquidity \\
R4 &Cash Flow/Total Debt &liquidity \\
R5 &Cash/Net Sales &liquidity \\
R6 &Cash/Total Assets &liquidity \\
R7 &Current Assets/Current Liabilities& liquidity \\
R8 &Current Assets/Net Sales &liquidity \\
R9 &Current Assets/Total Assets& liquidity \\
R10 &Current Liabilities/Equity &liquidity \\
R11 &Equity/Fixed Assets &solidity \\
R12 &Equity/Net Sales &solidity\\
R13 &Inventory/Net Sales &liquidity\\
R14 &Long Term Debt/Equity& solidity \\
R15 &Market Value of Equity/Book Value of Debt& solidity \\
R16 &Total Debt/Equity &solidity\\
R17 &Net Income/Total Assets& profitability \\
R18 &Net Quick Assets/Inventory& liquidity \\
R19 &Net Sales/Total Assets &profitability \\
R20 &Operating Income/Total Assets &profitability \\
R21 &Earnings Before Interest \& Tax/Total Interest Payments &liquidity \\
R22 &Quick Assets/Current Liabilities&liquidity \\
R23 &Quick Assets/Net Sales &liquidity \\
R24 &Quick Assets/Total Assets &liquidity \\
R25 &Rate of Return to Common Stock &profitability \\
R26 &Retained Earnings/Total Assets &profitability \\
R27 &Return on Stock &profitability\\
R28 &Total Debt/Total Assets& solidity \\
R29 &Working Capital/Net sales& liquidity \\
R30 &Working Capital/Equity &liquidity \\
R31 &Working Capital/Total Assets& liquidity \\
\hline
\end{tabular}
\label{tab1}
\end{table}
Also, the financial ratios $R31$ ($x_1$), $R26$ ($x_2$), $R21$ ($x_3$), $R15$ ($x_4$) and $R19$ ($x_5$) were identified as key variables for bankruptcy prediction and their weights are obtained by applying multivariate discriminate analysis. The final discriminate function obtained by Altman (say $Z_A$) is given by
\begin{equation}
\label{eqn2}
Z_A=1.2 \: x_1+1.4 \: x_2+3.3 \: x_3+0.6\: x_4+0.999\: x_5.
\end{equation}
The present condition of the firms may be assessed based on the $Z_A$ values as follows:
\begin{eqnarray}
\label{eqn4}
    Z_A=\left\{
                             \begin{array}{ll}
                             \hbox{Safe Zone}, & \hbox{$Z_A  \geq 2.99$;} \\ \nonumber
                            \hbox{Grey Zone}, & \hbox{$1.81 \leq Z_A< 2.99$;}\\ \nonumber
                            \hbox{Distress Zone}, &\hbox{$Z_A < 1.81$.} \nonumber
                             \end{array}
                           \right.
\end{eqnarray}
The companies in the safe zone may be considered financially healthy where as those in grey zone could go either way and if in the distress zone there is a greater risk that the company will go bankrupt within two years. \cite{34} refined the weights of the Z-scores by revisiting the Zeta analysis and obtained the discriminate function as
\begin{equation}
\label{eqn13}
Z_U = 0.72\: x_1 + 0.85 \: x_2 + 3.1 \: x_3 + 0.42 \: x_4 + x_5.
\end{equation}

In a linear model, the financial ratios $x_i$ would influence $Z$-score in a linear way. In the context of risk a change of a financial ratio by $1\%$ may not have the same influence on the score \citep{2,36,37}. Also, the studies in \citep{15,17} have found that financial ratios may be overstated due to accounting practices or manipulations. Therefore estimating nonlinear transformations for some of the independent variables would improve the bankruptcy predictions. The power transformations due to Box and Cox \citep{38} are a popular method for nonlinear transformations to improve the symmetry and normality of the model fit. However, these transformations are proposed only on positive values whereas the financial ratios could be negative. An alternative family of transformations is proposed in \citep{39} which may be applied even on negative values.

In the present work we propose a nonlinear mapping $x_i\mapsto f(x_i)$ of the form
\begin{eqnarray}
\label{eqn1}
    f(x_t)=\left\{
                             \begin{array}{ll}
                             -\ln (-x_t+1), & \hbox{ $x_t \leq 0$;} \\
                             \ln(x_t+1), & \hbox{ $x_t>0$,}
                             \end{array}    
						\right.                     
\end{eqnarray}
for transforming financial ratios ($x_t$) before deriving the Z-score. The loglinear models make the differences between large values less important and those between small values more important. They are employed in \citep{18,19} to assume proportionate growth in accounting variables that may be restricted to firm growth and are employed in \citep{20} for predicting the impending bankruptcy.

We first transform the financial ratios utilizing the nonlinear function given in Equation~\ref{eqn1} and then compute the new Z-score, $Z_M$ as
\begin{equation}
\label{eqn5}
Z_M=\lambda_1 \: f(x_1)+\lambda_2 \: f(x_2)+\lambda_3\: f(x_3)+\lambda_4 \:f(x_4)+\lambda_5 \:f(x_5)+\ldots+\lambda_t \:f(x_t),
\end{equation}
in which $\lambda_1 \ldots \lambda_t$ denote the parameters of the financial ratios $x_1 \ldots x_t$ respectively.
 
A more generalized form of $Z_M$ is given by 
  \begin{equation}
  \label{eqn3}
  Z_M=\sum_{k=1}^{t}\lambda_k \:f(x_k),
  \end{equation}
in which $x_{k}$ denotes the financial ratio for each $k=1,2,\ldots,t$ and $\lambda_{k}$ denotes the weight of the $x_{k}$. The weights $\lambda_k$ are estimated using multivariate discriminate analysis (MDA).

\section{A New Indexing Measure for Credit Rating}
\label{sec3}
To predict bankruptcy of a firm it is required to identify the bounds on Z-score that can be estimated through empirical studies. These bounds are not comparable and they vary from business to business and also on countries' economic situations. Therefore, there is a need to standardize the Z-score to a distribution before deriving useful indices for predicting the credit risk and  bankruptcy of the firms. The present procedure generalizes the bankruptcy prediction that can be utilized by all the companies around the world. In \citep{40} full credibility theory approach is used to estimate the parameters of the frequency (Poisson)  and severity (Pareto) distributions for low frequency, high impact operational risk losses exceeding some threshold for each risk cell. An extreme value theory that provided fundamentals needed for the statistical modelling has been applied in stock market indices \citep{22} to compute tail risk measures and extreme market events \citep{24}. A new set of assessment models for long-term credit risk is studied in \citep{44} which does not include stock prices and incorporates business cycles.

The final objective of any  risk model is to build the probability density function (PDF) of future losses in a portfolio. \cite{30} developed simplest model using Bernoulli-distributed events and Poisson distribution. Probability distributions such as Poisson and Gamma have been employed to analyze aggregate loss distributions associated with operational risk \citep{31,32,33}.  

Creditrisk$^{+}$ models assume that the risk factors are independent gamma distributed random variables with mean $1$ and variances $\sigma^{2}$ \citep{45}. Whereas, the P3 distribution with three parameters, location, shape and scale improves the goodness of fit to the data and can provide better estimates on the ratings. Moreover, Pearson family of distributions is employed in a wide range of applications such as financial time series modeling \citep{46}, distribution of stock returns  \citep{47}, flood risk modeling due to climate change \citep{48}.  This distribution can fit a wide range of shapes with positive or negative skewness including a good approximation to the normal distribution. This motivated us to develop a new methodology for credit risk rating using three parameter P3 distribution. In addition, our methodology is universal and can apply to a wide range of distributions that are popularly being employed in credit risk applications.

Presently, methods employed in credit risk applications use conventional moments such as mean, standard deviation, skewness and kurtosis to fit a distribution to the observations. These moments involve nonlinearities that are influenced by the presence of outliers and would result in over or underestimation of the credit ratings. Therefore, we propose an approach based on the method of linearized moments popularly known as L-moments where in the parameters of the distribution can be expressed in a linear form. These L-moments can be computed using probability weighted moments (PWM) presented in Section~\ref{sec3.1}. The parameters of the P3 distribution \citep{42} from L-moments can be computed using the procedure discussed in Section~\ref{sec3.2}. 

In our methodology we first propose to fit a P3 distribution to the $Z_M$ score by computing L-moments and the parameters of the distribution. An index is then computed by measuring the deviations of the data using parameters of the P3 distribution. The ratings are then obtained by classifying the index into intervals ranging from highest safety (AAA) to high risk (CCC) based on whether the value of the index is on positive or negative extreme of the distribution respectively. The details of the computations are presented in the Section~\ref{sec4}.

\subsection{Probability Weighted Moments and L-moments}
\label{sec3.1}
The L-moments are analogous to conventional central moments, but can be estimated by linear combinations of order statistics.  The L-moment estimates are found to be more robust compared to the conventional moments in the presence of outliers \citep{27,28,29}. The L-moments are less sensitive to the effects of sampling variability, and are used to characterize a wide range of distributions than the conventional moments. Practically, they are less subject to bias in estimation and they approximate their asymptotic normal distribution more closely. The parameters estimated through L-moments are more accurate than the maximum likelihood and least square estimates. The L-moment estimates using PWM has been used in applications such as floods~\citep{26}, drought~\citep{25} and financial risk \citep{23}.

The probability weighted moments are defined in terms of the cumulative distribution function $F(y)$ \citep{41}
\begin{equation}
\label{eqn6}
M_{p,r,s}=\int_{0}^{1}F^{-1}(y)^{p} F(y)^{r} (1-F(y))^{s} dF,
\end{equation}
in which $p$, $r$, and $s$ are positive integers, $F^{-1}(y)$ denotes the  inverse cumulative distribution function of the random variable $Y$. The term $M_{p,r,s}$ can now be used for describing the probability distribution. In a particular case where $p=1$, and $s=0$ the variable $y$ becomes linear and the moment $\beta_r$ is defined as
  \begin{equation}
  \label{eqn7}
\beta_r=M_{1,r,0}=\int_{0}^{1}F^{-1}(y) F(y)^{r} dF.
 \end{equation}
 The first three L-moments expressed as the linear combinations of the PWM as

 \begin{eqnarray}
 \label{eqn8}
 \theta_1=\beta_0, \\ \nonumber
 \theta_2=2 \beta_1-\beta_0, \\ \nonumber
 \theta_3=6 \beta_2-6 \beta_1+\beta_0,
 \end{eqnarray}
in which $\theta_1$ known as L-mean, is a measure of central tendency and $\theta_2$ known as L-standard deviation is a measure of dispersion. The ratios of L-moments are defined as
\begin{eqnarray}
 \label{eqnlratio}
 \tau_2=\theta_2 / \theta_1,\\ \nonumber
 \tau_3=\theta_3/ \theta_2, 
 \end{eqnarray}
in which $\tau_2$ is termed as L-coefficient of variation and $\tau_3$ is known as L-skewness and they are employed in estimating the parameters of the P3 distribution.

\subsection{Pearson Type 3 (P3) Distribution}
\label{sec3.2}

In particular, the P3 probability density function $g$ of the random variable $\Xi$ is defined as
\begin{equation}
\label{eqn9}
g(\xi)=\frac{\left|\alpha\right|}{\Gamma(\eta)} [\alpha (\xi-c)]^{\eta-1} e^{-\alpha\left(\xi-c)\right.},
\end{equation}
in which $c$, $\alpha$ and $\eta$ are location, scale and shape parameters of the distribution respectively. When the parameter $\alpha >0$, $\xi$ has positive skewness leading to $c \leq \xi \leq +\infty$ and when $\alpha <0$, $\xi$ has negative skewness leading to $-\infty  \leq \xi \leq c $. Hence, $c$ is a lower bound for positively skewed and an upper bound for negatively skewed P3 random variable $\Xi$.

The parameters $c$, $\alpha$ and $\eta$ of P3 distribution are related to the L-moments as 
\begin{eqnarray}
    \eta &=&\left\{
                             \begin{array}{ll}
                               \frac{1+0.2906 \: \delta}{\delta+0.1882 \: \delta^2+0.0442 \: \delta^3}, & \hbox{$0 < \tau_3 < 0.3333$;} \\
                             \frac{0.36067 \: \zeta - 0.5967 \: \zeta^2+0.2536\: \zeta^3}{1-2.78861 \zeta+2.56096 \zeta^2 - 0.77045 \: \zeta^3} , & \hbox{$0.3333 \leq \tau_3 <1$}
													 \nonumber 	                    
										\end{array}	
\right.										
\end{eqnarray}

\begin{eqnarray}
\label{eqnparamp3}
\begin{array}{lll}              
\alpha & = &\sqrt{\pi} \: \theta_2 \: e^{\left(\Gamma(\eta)-\Gamma(\eta+0.5)\right.)} ,\\ 

c & = & \theta_1 - \left(\alpha \: \eta \right),
\end{array}	
\end{eqnarray}
in which $\delta = 3 \: \pi \tau_3^2$ and $\zeta =1-\tau_3$.

\section{Methodology}
\label{sec4}
\begin{figure}[ht]
	\includegraphics[height=90mm,width=115mm]{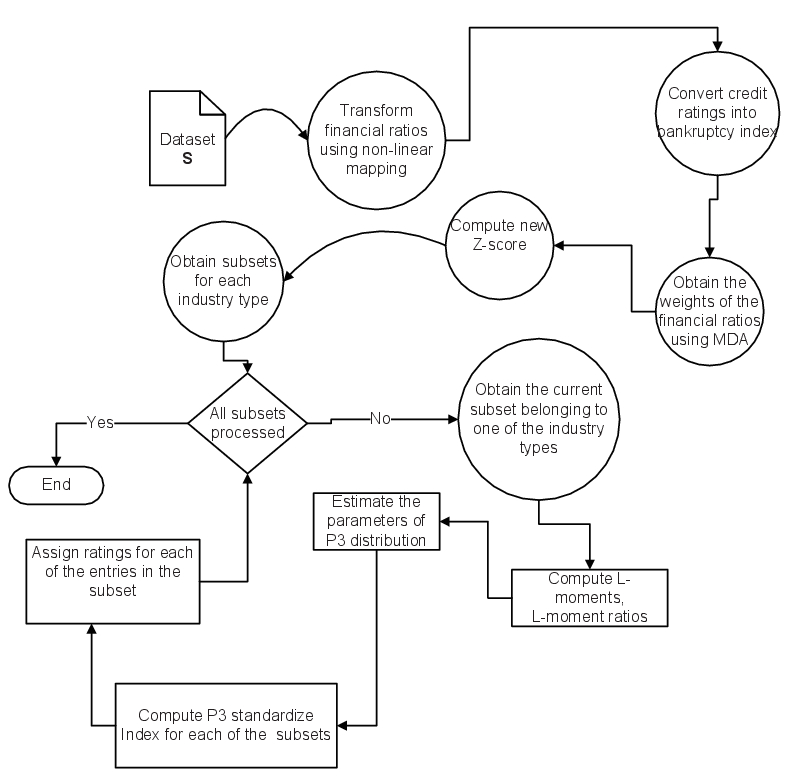}\\
  \caption{Methodology for computing the new index measure}\label{fig1}
\end{figure}

The steps involved in the computation of the index are shown in Fig. \ref{fig1}. The dataset consisting of the financial ratios is first transformed into new variables using the nonlinear function proposed in Equation \ref{eqn3}. In the next step we convert the credit ratings given in the dataset into a binary variable $b_{\phi}$ henceforth known as bankruptcy index as follows. Define

\begin{eqnarray}
\label{eqn10}
    b_{\phi}=\left\{
                             \begin{array}{ll}
                              1, & \hbox{$\forall$ $R_{\phi}$ $\in$ $\{B, BB, BBB, CCC\}$;} \\ 
                              0, & \hbox{$\forall$ $R_{\phi}$ $\in$ $\{A,AA,AAA\}$,} 
                             \end{array}
                           \right.
\end{eqnarray}
in which $R_{\phi}$ is the credit rating of the record $\phi$ in the data set of $m$ records,  i.e., $\phi$ takes values from $1, \ldots, m$. As per CRISIL the credit ratings $AAA$ denotes highest safety, $AA$ denotes high safety, $A$ denotes adequate safety, $BBB$ denotes moderate safety, $BB$ denotes moderate risk, $B$ denotes high risk and $CCC$ denotes very high risk. Clearly, from the Equation \ref{eqn10} one can infer that $b_{\phi} =1 \Rightarrow $ bankruptcy or high risk category and $b_{\phi}=0 \Rightarrow $ non-bankruptcy or high safety category.

Subsequently, the weights $\lambda_1,\ldots,\lambda_t$ of the transformed financial ratios are estimated using MDA with $b_{\phi}$ as dependent variable and $f(x_1),f(x_2),\ldots,f(x_t)$ as independent variables. 

\begin{eqnarray}
\label{eqn14}
b_{1}= \lambda_1 \: f(x_{1})_1+\lambda_2 \: f(x_{2})_1+,\ldots,+\lambda_t \:f(x_{t})_1, \nonumber \\
\vdots \nonumber \\
b_{\phi}=\lambda_1 \: f(x_{1})_\phi+\lambda_2 \: f(x_{2})_\phi+,\ldots,+\lambda_t \:f(x_{t})_\phi, \nonumber \\
\vdots \nonumber \\
b_{m}=\lambda_1 \: f(x_{1})_m+\lambda_2 \: f(x_{2})_m+,\ldots,+\lambda_t \:f(x_{t})_m, 
\end{eqnarray}
in which the variables $f(x_1),f(x_2),\ldots,f(x_t)$ denote the financial ratios after the application of the function $f$ on the financial ratios $x_1,x_2,\ldots,x_t$ respectively as defined in Equation \ref{eqn1}. The weights obtained from the Equation \ref{eqn14} are substituted in Equation \ref{eqn5} to obtain the score $Z_M$.

The new Z-score $Z_M$ is then split into subsets $Z_M=\{Z_{1,\{1,\ldots,j\}},Z_{2,\{1,\ldots,j\}},\\ \ldots,Z_{i,\{1,\ldots,j\}} \}$ where in $j$ denotes the year of observation for the $i^{th}$ industry type. For each of these subsets the PWM are computed using the Equation \ref{eqn7}. The L-moments $\theta_1, \theta_2, \theta_3$, L-moment ratios $\tau_2$ and $\tau_3$ are computed using Equations \ref{eqn8}, \ref{eqnlratio} respectively. The parameters $c, \eta, \alpha$ of P3 distribution are obtained from L-moments and L-moment ratio ($\tau_3$) by employing Equation \ref{eqnparamp3}. 

In the next step we standardize the dataset $Z_{M_{i,j}}$ with respect to the origin (parameter $c$) of the P3 distribution as $v_{i,j}=(Z_{M_{i,j}}-c)/ \alpha.$ The new index $H_{i,j}$ is then obtained as $H_{i,j}=\left((v_{i,j}/\eta)^{0.33}+1/(9\eta)-1 \right) \left(9\eta \right)^{0.5}$. Based on the index the credit ratings are assigned to the dataset. The details of the procedure are presented in the Algorithm~\ref{alg1}.

\begin{algorithm}
\caption{Algorithm for Computing Ratings based on the novel Credit Risk Index}
\label{chap9:PRA}
\begin{algorithmic}
\Require
\begin{enumerate}
   \item Data set $\mathbf{S}(m,n)$ where $m$ and $n$ denotes the number of rows and number of columns respectively. The attributes consist of financial ratios $x_1(m,1)$, \ldots, $x_t(m,1)$ and credit ratings $\mathbf{R}(m,1)$ as per CRISIL. Clearly the number of columns in the dataset $\mathbf{S}$ is $n=t+1$.
	\item  A set $\mathbf{E}(1,m)$ consisting of the type of industry type as in \cite{43} numbered from $1,\ldots,12$ to which each row in the data set $\mathbf{S}$ belongs.
	\item A set $\mathbf{Y}(1,m)$ consisting of the year in which the ratings are observed for each row in the dataset $\mathbf{S}$.
\end{enumerate}
\Ensure
\begin{enumerate}
            \item  Ratings $\mathbf{W}(1,m)$ for each row in the dataset $\mathbf{S}$.
\end{enumerate}
   \textbf{Algorithm}
   \begin{enumerate}
        \item Transform the financial ratios given in the dataset $\mathbf{S}$ using the nonlinear transformation function $f$ given in Equation \ref{eqn1} to obtain a set with columns $f(x_1),\ldots,f(x_t)$. Designate the set as $\mathbf{D}(m,t)$. The set $\mathbf{D}$ has $m$ records and $t$ attributes.  
				\item Convert the credit ratings in column $R$ in the dataset $\mathbf{S}$ to a bankruptcy index variable $b$ using the transformation in Equation \ref{eqn10}. We designate this set as $\mathbf{\Theta}$. Obtain the set $\mathbf{\tilde{\Theta}}=\mathbf{\Theta} \cup \mathbf{D}$. The set  $\mathbf{\tilde{\Theta}}(m,\hat{n})$ has $m$ rows and $\hat{n}= t+1$ columns.
			\item Obtain the weights $\lambda_1, \ldots, \lambda_t$ in Equation \ref{eqn14} by employing MDA with column $b$ of the set $\mathbf{\tilde{\Theta}}$ as dependent variable and the columns $f(x_1),\ldots,f(x_t)$ of the set $\mathbf{\tilde{\Theta}}$ as independent variables.
        \item Obtain the new Z-score using the computations given in Equation \ref{eqn5} for all the records in the dataset $\mathbf{\tilde{\Theta}}$ with the weights obtained in Step $3$. Designate the set as $\mathbf{Z_M}(1,m)$. Obtain the set $\mathbf{\Omega}=\mathbf{E} \cup \mathbf{Y} \cup \mathbf{\tilde{\Theta}} \cup  \mathbf{Z_M}$. Clearly the set $\mathbf{\Omega}(m,\tilde{n})$ has $m$ rows and $\tilde{n}=\hat{n}+3$ columns.
				 \item For each type of industry $i$ in column $\mathbf{E}$ of set $\mathbf{\Omega}$
				\begin{enumerate}
        \item identify and collect all records in $\mathbf{\Omega}$ belonging to a particular type of industry $E(i)$. The subset is designated as $\mathbf{Q}$ i.e $\mathbf{Q}_{\tilde{m},\tilde{n}}=\{\mathbf{\Omega}_{m,\tilde{n}} : m \in E(i)\}$. Clearly $\mathbf{Q} \subseteq \mathbf{\Omega}$ and $\tilde{m} \leq m$.  
          \item obtain the PWM using the $\mathbf{Z_M}$ column in set $\mathbf{Q}$ and then compute L-moments and their ratios using Equations \ref{eqn8}, \ref{eqnlratio}.
        \item compute the parameters $c, \alpha, \eta$ of P3 distribution using Equation \ref{eqnparamp3}. 
				\item for each year $j$ in column $\mathbf{Y}$ of set $\mathbf{Q}$
        \begin{enumerate}
        \label{step3}
				\item compute the quantity $v_{i,j}=(Z_{M_{i,j}}-c)/ \alpha$.
				\item obtain the index $H_{i,j}=\left((v_{i,j}/\eta)^{0.33}+1/(9\eta)-1 \right) \left(9\eta \right)^{0.5}$.
				\item obtain the row identification ($d$) corresponding to the $i^{th}$ industry and $j^{th}$ year i.e $d=Index(i,j)$ would return the row number in dataset $\mathbf{Q}$ for which the rating is being computed.
        \item compute the rating as defined by
        \begin{eqnarray}
            \label{eqn12}
                  \mathbf{W}_{d}=\left\{
                             \begin{array}{ll}
                             AAA, & \hbox{$H_{i,j}>2.0$;} \\
                           AA, & \hbox{$1.5< H_{i,j} \leq 2.0$;}\\
                           A, & \hbox{$0< H_{i,j}\leq 1.5$;} \\
                             BBB, & \hbox{$-1.0< H_{i,j}\leq 0.0$;}\\
                             BB, & \hbox{$-1.5 < H_{i,j} \leq -1.0$;} \\
                             B, & \hbox{$-2.0< H_{i,j} \leq -1.5$;}\\
                             CCC, & \hbox{$H_{i,j} \leq -2.0$,}\\
                             \end{array}
                           \right.
        \end{eqnarray}
				\noindent in which $\mathbf{W}$ denotes the set containing the new ratings.
       \end{enumerate}
			   \item initialize the set $\mathbf{Q}$ i.e Let $\mathbf{Q}=\{\emptyset\}$.
       \end{enumerate}
		\item RETURN $\mathbf{W}$.
    \item END.
   \end{enumerate}
\end{algorithmic}
\label{alg1}
\end{algorithm}

\newpage

\subsection{Toy Example} 
To illustrate our methodology, we utilize a toy dataset given in Table \ref{tab9} with $m=10$ records, financial ratios in columns $x_1, \ldots, x_5$ (say $t=5$), and the credit ratings in column $\mathbf{R}$ as per CRISIL. Clearly the total number of attributes $n=t+1=5+1=6$. The dataset is designated as $\mathbf{S}(m,n).$
\begin{table}[H]
\caption{Toy dataset}
\centering
\begin{tabular}{llllll}
\hline\hline
$x_1$	&	$x_2$	&	$x_3$	&	$x_4$	&	$x_5$	&	$R$	\\ [0.5ex]
\hline
	0.121	&	0.263	&	0.046	&	1.219	&	0.286	&	BBB	\\
-0.046	&	-0.164	&	0.027	&	0.218	&	0.103	&	B	\\
	0.481	&	0.696	&	0.099	&	3.969	&	0.532	&	AAA	\\
	0.351	&	0.238	&	0.07	&	1.023	&	0.237	&	BBB	\\
	0.217	&	0.326	&	0.045	&	2.522	&	0.295	&	AA	\\
	0.105	&	0.236	&	0.053	&	1.566	&	0.216	&	BBB	\\
	0.078	&	0.157	&	0.041	&	1.402	&	0.335	&	BBB	\\
	0.189	&	0.437	&	0.059	&	5.043	&	0.452	&	AAA	\\
	0.043	&	-0.047	&	0.041	&	0.287	&	0.114	&	B	\\
	0.17	&	0.702	&	0.089	&	23.002	&	1.183	&	AAA	\\
\hline
\end{tabular}
\label{tab9}
\end{table}
The dataset can as well be written as 
\begin{align} \textbf{S} &=&\{(0.121, 0.263, 0.046, 1.219, 0.286, BBB), (-0.046, -0.164, 0.027, 0.218, 0.103, B), (0.481, 0.696, 0.099, 3.969, 0.532, AAA), (0.351, 0.238, 0.07, 1.023, 0.237, BBB), (0.217, 0.326, 0.045, 2.522, 0.295, AA), (0.105, 0.236, 0.053,	1.566, 0.216, BBB), (0.078, 0.157, 0.041, 1.402, 0.335, BBB), (0.189, 0.437, 0.059,	5.043, 0.452,	AAA), (0.043, -0.047, 0.041, 0.287, 0.114, B), (0.17, 0.702, 0.089, 23.002, 1.183, AAA)\}.
\end{align}
Clearly the cardinality of above set is $|\mathbf{S}|=10$. We now present the steps involved in Algorithm \ref{alg1} for computing the credit ratings. We consider sets $\mathbf{E}(1,m)=\{1,1,1,1,1,1,1,1,1,1\}$ and $\mathbf{Y}(1,m)=\{1,2,3,4,5,6,7,8,9,10\}$ denoting the type of industry and the year of observation respectively, for each of the records in $\mathbf{S}$. Clearly the cardinality of the sets $\mathbf{E}$ and $\mathbf{Y}$ is equal to the number of rows in $\mathbf{S}$ i.e., $|\mathbf{E}| = |\mathbf{Y}|=10$. A log-transformation function $f$ given in Equation \ref{eqn1} is applied to each of the financial ratios (i.e for column $x_1$ of record $1$ we have $f(x_{{1}_1})=f(0.121)=\ln(0.121+1)=0.114$). We designate this set as $\mathbf{D}$. Clearly the elements of this set are \begin{align} \textbf{D}&=&\{(0.114, 0.233, 0.045, 0.797, 0.252), (-0.045, -0.152, 0.027, 0.197, 0.098), (0.393, 0.528, 0.094, 1.603, 0.427), (0.301, 0.213, 0.068, 0.705, 0.213),  (0.196, 0.282, 0.044, 1.259, 0.259), (0.1, 0.212, 0.052, 0.942, 0.196), (0.075, 0.146, 0.04, 0.876, 0.289), (0.173, 0.363, 0.057, 1.799, 0.373), (0.042, -0.046, 0.04, 0.252, 0.108), (0.157, 0.532, 0.085, 3.178,	0.781)\}.\end{align} We obtain the bankruptcy index $b$ by applying Equation \ref{eqn10} to the column $R$ in the Table \ref{tab9} (i.e. record $1$, $R_{1} \in BBB \Rightarrow b_{1}=1$). We designate this set as $\mathbf{\Theta}$ and the elements of this set are $\{ 1, 1, 0, 1, 0, 1, 1, 0, 1, 0\}.$ We then obtain a set $\mathbf{\tilde{\Theta}}= \mathbf{D} \cup \mathbf{\Theta}$. The elements of this set are $\mathbf{\tilde{\Theta}}$ are \begin{align} &=&\{(0.114, 0.233, 0.045, 0.797, 0.252, 1), (-0.045, -0.152, 0.027, 0.197, 0.098, 1), (0.393, 0.528, 0.094, 1.603, 0.427, 0), (0.301, 0.213, 0.068, 0.705, 0.213, 1), (0.196, 0.282, 0.044, 1.259, 0.259, 0), (0.100, 0.212, 0.052, 0.942, 0.196, 1), (0.075, 0.146, 0.04, 0.876, 0.289, 1), (0.173, 0.363, 0.057, 1.799, 0.373, 0), (0.042, -0.046, 0.04, 0.252, 0.108, 1), (0.157, 0.532, 0.085, 3.178,	0.781, 0)\}.\end{align} 

We obtain the set of equations between the bankruptcy index $b$ and the financial ratios $x_1, \ldots, x_5$ as given in Equation \ref{eqn14}.

\begin{eqnarray}
\label{eqn15}
b_{1}=1=0.114 \: \lambda_1 +0.233 \: \lambda_2 +0.045 \: \lambda_3 +0.797 \: \lambda_4  + 0.252 \: \lambda_5,\nonumber \\ 
b_{2}=1=-0.045 \: \lambda_1 -0.152 \: \lambda_2 + 0.027 \: \lambda_3  +0.197 \: \lambda_4  + 0.098 \: \lambda_5, \nonumber \\ 
\vdots \nonumber \\ 
b_{10}=0=0.157 \: \lambda_1  +0.532 \: \lambda_2  +0.085 \: \lambda_3  + 0.178 \: \lambda_4  + 0.781 \: \lambda_5.
\end{eqnarray}

By employing MDA on Equation \ref{eqn15} we obtain the weights as $\lambda_1=1.841$, $\lambda_2=-0.856$, $\lambda_3=-1.087$, $\lambda_4=3.390$, $\lambda_5=-1.649.$

The score $Z_M$ is computed using Equation \ref{eqn5} as 
\begin{eqnarray}
\label{eqn16}
Z_{M_{1}}= 1.841\times 0.114-0.233 \times 0.856- 0.045 \times 1.087 
						\nonumber \\	+0.797 \times 3.390 - 0.252 \times 1.649=2.249,  \nonumber \\			
\vdots \nonumber \\
Z_{M_{10}}= 0.157 \times 0.114  - 0.532 \times 0.856  - 0.085 \times 1.087 
              \nonumber \\ +0.178 \times 3.390 - 0.781 \times 1.649=9.228.
\end{eqnarray}

The computed $Z_M$ scores obtained for the years $j=1,\ldots,10$ using Equation \ref{eqn16} is given as$Z_M=\{2.249,	0.525,	4.900,	2.335,	3.914,	2.818,	2.464,	5.429,	0.750,	9.228\}.$ We then obtain the dataset $\mathbf{\Omega}=\mathbf{E} \cup \mathbf{Y} \cup \mathbf{\tilde{\Theta}} \cup  \mathbf{Z_M}$ as given in Table \ref{tab10}.
\begin{table}[H]
\caption{Data set obtained after applying function $f$ in Equation \ref{eqn1} and bankruptcy index $b$ in Equation \ref{eqn10} and new Z-score $Z_M$ using data in Table \ref{tab9}}
\centering
\begin{tabular}{cclllllll}
\hline\hline
$E$ & $Y$	&	$f(x_1)$	&	$f(x_2)$	&	$f(x_3)$	&	$f(x_4)$	&	$f(x_5)$	&	$b$	& $Z_M$\\
\hline
1	&	1	&	0.114	&	0.233	&	0.045	&	0.797	&	0.252	&	1	& 2.249\\
1	&	2	&	-0.045	&	-0.152	&	0.027	&	0.197	&	0.098	&	1	&0.525\\
1	&	3	&	0.393	&	0.528	&	0.094	&	1.603	&	0.427	&	0	&4.900\\
1	&	4	&	0.301	&	0.213	&	0.068	&	0.705	&	0.213	&	1	&2.335\\
1	&	5	&	0.196	&	0.282	&	0.044	&	1.259	&	0.259	&	0	&3.914\\
1	&	6	&	0.100	&	0.212	&	0.052	&	0.942	&	0.196	&	1	&2.818\\
1	&	7	&	0.075	&	0.146	&	0.040	&	0.876	&	0.289	&	1	&2.464\\
1	&	8	&	0.173	&	0.363	&	0.057	&	1.799	&	0.373	&	0	&5.429\\
1	&	9	&	0.042	&	-0.046	&	0.040	&	0.252	&	0.108	&	1&0.750	\\
1	&	10	&	0.157	&	0.532	&	0.085	&	3.178	&	0.781	&	0	&9.228\\

\hline
\end{tabular}
\label{tab10}
\end{table}

The PWM are computed using Equation \ref{eqn7} to obtain the values $\beta_0=3.461$, $\beta_1=2.449$, and $\beta_2=1.939.$

The L-moments and their ratios are computed from $\beta_0,\beta_1,\beta_2$ using Equation \ref{eqn7}, \ref{eqnlratio} as 
\begin{eqnarray}
\label{eqn17}
\theta_1=\beta_0=3.461, \nonumber \\
\theta_2=2 \times 2.449-3.461=1.437, \nonumber \\
\theta_3= 6 \times 1.939 - 6 \times 2.449+3.461=0.401, \nonumber \\
\tau_2=\theta_2/\theta_1=1.437/3.461=0.415,\nonumber \\
\tau_3=\theta_3/\theta_2=0.401/1.437=0.279.
\end{eqnarray}

The parameters of P3 distribution is obtained by substituting from L-moments obtained in Equation \ref{eqn17} in Equation \ref{eqnparamp3}. Since, $\tau_3 < 0.333$ we apply

\begin{eqnarray}
\label{eqn18}
\delta=3 \times 3.146 \times (0.279)^2= 0.7202, \nonumber \\
\eta=\frac{(1 + 0.2906 \times 0.7202)}{(0.7202 + 0.1882 \times (0.7202)^2 + 0.0442 \times (0.7202)^3)} \nonumber \\
=\frac{(1+0.2093)}{(0.7202+0.0976+0.0165)}=1.449, \nonumber \\ 
\alpha=\sqrt{3.1416} \times 1.437 \times e^{(\Gamma(1.449)-\Gamma(1.449+0.5))} \nonumber \\
      = 1.7725 \times 1.437 \times e^{(-0.1214-(-0.0205))} \nonumber \\
			= 2.5488 \times e^{-0.1009}= 2.3042,\nonumber \\
c=3.461- 1.449 \times 2.3042= 3.461-3.3398=0.121.
\end{eqnarray}

To compute the Index $H$ for $i=1$ and $j=1$ we first compute 
\begin{eqnarray}
\label{eqn19}
v_{1,1}=(Z_{M_{1,1}}^\mathsf{T}-c)/\alpha=(2.249-0.121) /2.3042 \nonumber \\
=2.1273/2.3042=0.9232, \nonumber \\
H_{1,1}=\left((v_{1,1}/1.449)^{0.33}+1/(9\times1.449)-1 \right) \left(9\times1.449 \right)^{0.5}  \nonumber \\
 =\left((0.9232/1.449)^{0.33}+1/(9\times1.449)-1 \right) \times \left(9\times1.449 \right)^{0.5}  \nonumber \\
=\left(0.8604+0.0767-1 \right) \times 3.6118 \nonumber \\
=-0.0629 \times 3.6118= -0.2272.
\end{eqnarray}

The remaining indices of $H$ for $i=1$ and $j=2,\ldots,10$ can be obtained using the steps in Equation \ref{eqn19} as $H_{i,j}$\begin{align}&=&\{-1.549, 0.735, -0.186, 0.433, 0.028, -0.126, 0.880, -1.265, 1.711\}. \end{align} To compute $\mathbf{W}_{1}$ we apply Equation \ref{eqn12} on $H_{1,1}= -0.2272$ and find that it falls in the range $\hbox{$-1.0< -0.2272\leq 0.0$}$, hence the rating $BBB$ is assigned to $\mathbf{W}_{1}.$

Similarly, the credit ratings for $i=1$ and $j=2,\ldots,10$ can be obtained using Equation \ref{eqn12} on values in $H_{i,j}$ as $ \mathbf{W}_{2,\ldots,10}$\begin{align}&=&\{B, A, BBB, A, A, BBB, A, BB, AA\}. \end{align}


\section{Experiments and Results}
\label{sec5}
In this section we present the experiments conducted on the dataset and a comparison of our results with those of the earlier studies. A time series dataset \cite{44} consisting of $3932$ records with seven attributes as given in Table \ref{tab2} is considered in our analysis. We have considered five financial ratios namely (i) working capital/total assets (WC\_TA), (ii) retained earnings/total assets (RE\_TA), (iii) earnings before interest and taxes/total assets (EBIT\_TA), (iv) market value of equity/book value of the total debit (MVE\_BVTD) and (v) sales/total assets (S\_TA) for the analysis. There are $2392$ bankrupt cases with credit ratings from $BBB$ to $CCC$ and $1540$ nonbankrupt cases with credit ratings $A$ to $AAA.$

\begin{table}[H]
\caption{Description of the dataset}
\centering
\begin{tabular}{clp{35mm}l}
\hline\hline
Sno&Attribute&Description&Type\\ [0.5ex]
\hline
1&WC\_TA	& working captial/total assets& Real \\
2&RE\_TA	& retained earnings/total assets&Real\\
3&EBIT\_TA&earnings before interest and taxes/total assets&Real	\\
4&MVE\_BVTD&market value of equity/book value of total debit&Real	\\
5&S\_TA	&sales/total assests&Real\\
6&Industry& 1 to 12& Categorical	\\
7&Rating& A, AA, AAA, B, BB, BBB, CCC&Categorical\\
\hline
\end{tabular}
\label{tab2}
\end{table}

The dataset consists of the credit ratings belonging to $12$ different industries with seven ratings ranging from highest safety (AAA) to very high risk (CCC).

\subsection{Results}
In this section we present the results of our study on the dataset. A comparison of skewness and the kurtosis of the original and transformed financial ratios are shown in Table \ref{tab8}.
\begin{table}[ht]
	\caption{Comparison of skewness and kurtosis of old and transformed financial ratios}
\centering
\begin{tabular}{p{20mm}llp{9mm}lp{9mm}}
\hline\hline
Financial&Skewness &Skewness  & Kurtosis  & Kurtosis\\ 
ratio& old& transformed &  old & transformed\\ 
\hline
WC\_TA	&-1.152&	-0.458	&17.944	&4.637 \\
RE\_TA	&-2.476&	-1.591	&17.181	&6.462 \\
EBIT\_TA	&-4.665	&-3.760	&74.310	&51.487 \\
MVE\_BVTD	&12.992	&1.415	&269.574	&3.357 \\
S\_TA	&9.160	&2.129	&206.135	&12.598 \\
\hline
\end{tabular}
\label{tab8}
\end{table}

From Table \ref{tab8} we can infer that the log transformation has reduced the skewness and kurtosis of the original variables, thereby improving the normality of the financial ratios.

%
%
%
%

We then estimate the weights of the loglinear model by performing MDA analysis between the bankruptcy index obtained by applying Equation \ref{eqn10} and log transformed financial ratios. We have obtained the parameters given in the Equation \ref{eqn3} as $\lambda_1=0.375$, $\lambda_2=0.028$, $\lambda_3=-0.316$, $\lambda_4=1.126$, and $\lambda_5=-0.236$.

Altman's Z-score ($Z_A$) and the revised Z-score ($Z_U$) is computed as given in Equations \ref{eqn2} and \ref{eqn13}. The $Z_M$ score of the transformed variables is  computed and comparison of the descriptive statistics among the $Z_M$,  original Z-score and revised Z-score score is shown in Table \ref{tab3}.

\begin{table}[ht]
\caption{Descriptive Statistics among the new score ($Z_M$), Altman's Z-score ($Z_A$) and revised Z-score ($Z_U$)}
\centering
\begin{tabular}{p{25mm}llp{9mm}lp{9mm}}
\hline\hline
Scoring&Mean&Standard &Quantile&Median& Quantile \\ 
Method&&deviation&(25\%)&& (75\%) \\
\hline
$Z_M$&	1.007&	0.686 &0.529&0.883 &1.374 \\
$Z_A$&	2.172&	3.060&0.926&1.626&2.679 \\
$Z_U$&1.609&2.178&0.714&1.207&1.973 \\
\hline
\end{tabular}
\label{tab3}
\end{table}
We performed F-Test on the standard deviation of the $Z_M$ and other Z-score methods. The estimated $F$ value ($F_{cal}$) is $30.664$ and the tabulated $F$ value ($F_{tab}$) is $39.863$ at $0.01$ significance. Since $F_{cal}$ is less than $F_{tab}$ we accept the null hypothesis that the two standard deviations are equal. The original $Z_A$ and $Z_M$ are found to have good correlation with Spearman $rho$ coefficient $0.963$ significant at $0.01$ ($2$ tailed) level (see Table \ref{tab4}).
\begin{table}[ht]
\caption{Correlation coefficients between $Z_M$ and $Z_A$}
\centering
\begin{tabular}{l l l}
\hline\hline
&$Z_M$&$Z_A$ \\ [0.5ex]
\hline
$Z_M$&1.0&0.963\\
$Z_A$ &0.963&1.0\\
\hline
\end{tabular}
\label{tab4}
\end{table}

The estimates of the parameters of the P3 distribution obtained using the methodology described in Section \ref{sec4} are shown in Table \ref{tab5} for each of the industry types.

\begin{table}[H]
\caption{P3 distribution parameter estimates using probability weighted moments}
\centering
\begin{tabular}{clll}
\hline\hline
Industry Type&Location ($\epsilon$)&Shape ($\beta$)&Scale ($\alpha$) \\ [0.5ex]
\hline
1&-0.120815&0.434659&2.588683 \\
2&-0.141720&0.387980&3.053867\\
3&-0.466324&0.279429&5.150730\\
4&-0.168323&0.422125&2.898292\\
5&-0.157205&0.466022&2.546405\\
6&-0.469409&0.290401&5.071536\\
7&-0.640034&0.261942&6.265737\\
8&-0.279976&0.341867&3.639809\\
9&-0.062825&0.456268&2.352211\\
10&-0.117848&0.396729&2.796561\\
11&-0.303738&0.344976&3.710743\\
12&-0.186726&0.382682&3.192083\\
\hline
\end{tabular}
\label{tab5}
\end{table}

The credit ratings are computed after obtaining the standardized index of the data set using the P3 distribution parameters. The credit ratings are then converted to bankruptcy index using the Equation \ref{eqn10}. A binary logistic regression, classification is carried out with the bankruptcy index as dependent variable and the $Z_A$, $Z_M$ or the $Z_U$ scores as an independent variable as given in Equation \ref{eqnmdaz}
\begin{eqnarray}
\label{eqnmdaz}
\Lambda_{a}=\nu_a+ \nu_1 \: Z_{A}  +\epsilon_{1},  \nonumber \\
\Lambda_{m}=\nu_m+ \nu_2 \: Z_{M} +\epsilon_{2}, \\
\Lambda_{u}=\nu_u+ \nu_3 \: Z_{U} +\epsilon_{3}, \nonumber
\end{eqnarray}
\noindent in which $\Lambda_{a}$, $\Lambda_{m}$, $\Lambda_{u}$ denotes the bankruptcy index of the Z-scores $Z_A$, $Z_M$ and $Z_U$ respectively.

The parameters obtained from the three models are shown in Table \ref{tab6}.
\begin{table}[H]
\caption{Comparison of $Z_A$, $Z_U$ and the new score $Z_M$ for bankruptcy prediction using Wald statistics}
\centering
\begin{tabular}{clll}
\hline\hline
Parameter&$Z_A$&$Z_M$&$Z_U$ \\ [0.5ex]
\hline
$\nu_a$&11.016(594.168)	& - &-\\
$\nu_m$ &- & 77.15(106.313) &- \\
$\nu_u$ &- & -&11.956(574.677)\\
$\nu_1$&-5.423(572.703)& -&-\\
$\nu_2$&-&-78.534(106.569)&- \\
$\nu_3$&-&-&-7.993(555.379)\\
\hline
\end{tabular}
\label{tab6}
\end{table}
The estimated coefficients of $Z_A$ and $Z_M$ are both negative and statically significant at $0.01\%$ indicating that both measures are useful in predicting bankruptcy risk and lower the score, the higher the risk of bankruptcy. The coefficient of $Z_M$ is far lower than $Z_A$ indicating the fact that the predictive power of $Z_M$ is far better than Altman's $Z_A$ score and revised Altman's $Z_U$ score. 

A hold-out classification between $Z_M$, $Z_A$, $Z_U$ and bankruptcy index $b$ is carried out using MDA and the prediction accuracy of the proposed methodology is found to be $98.6\%$ which is higher by $5\%$ than any of the models proposed by Altman. This confirms that the proposed methodology is universal and serves as a generalized tool that can improve the estimations of the existing methods/procedures in vogue and can predict the bankruptcy risk in an effective manner.



Discriminate analysis on the data set generated using the new transformation $Z_M$ has resulted in an accuracy of $93.7\%$ in cross validated grouped cases correctly classified where as Altman's Z-score $Z_A$ has resulted only in an accuracy of $87.4\%$.

MDA is carried out on $Z_M$ score and the credit ratings obtained from P3 and Pareto distributions. The proposed  method with P3 distribution resulted in an accuracy $92.2\%$ whereas the model with the Pareto distribution has resulted in only $80\%$ accuracy.

To understand the sensitivity on the choice of thresholds for different ratings in predicting bankruptcy, we first construct a classification matrix or accuracy matrix (Table \ref{tabclassmatx}) based on the number of agreements and disagreements between the predicted group membership (estimated from the model) and the actual group membership (as present in the dataset) of bankruptcy for the thresholds given in Equation \ref{eqn12}. 
\begin{table}[H]
\caption{Classification matrix for the thresholds given in Equation \ref{eqn12}}

\begin{tabular}{|l||l|l|}
\hline
 &\multicolumn{2}{l|}{Predicted group membership}\\
\cline{2-3}
 Actual group membership&Bankrupt& Non-Bankrupt\\
\hline\hline
Bankrupt&1966 ($N_1$)&426 ($M_1$)\\
 Non-Bankrupt&14 ($M_2$)&1526 ($N_2$)\\
\hline
\end{tabular}
\label{tabclassmatx}
\end{table}
The actual group membership is equivalent to the a priori grouping and the predicted group refers to the cases wherein the proposed methodology attempts to classify them correctly. In the Table \ref{tabclassmatx} $N_1$, $N_2$ denotes the correct classifications (Hits) and $M_1$, $M_2$ denotes the misclassifications (Misses). $N_1$ ($1966$) gives the number of cases of actual bankruptcy correctly classified as bankrupt by the proposed method. $M_1$ ($426$) is the Type-I error that gives the number of cases wherein the actual group membership is bankrupt whereas the proposed model misclassified them as non-bankrupt. $M_2$ ($14$) is the Type-II error,  that denotes the number of actual cases belonging to non-bankrupt group misclassified as bankrupt by the proposed model. $N_2$ ($1526$) are the number of cases wherein the proposed model correctly labels the actual cases as non-bankrupt. 

The accuracy of the proposed methodology is computed as $(N_1+N_2)/(N_1+N_2+M_1+M_2)=(1966+1526)/(1966+1526+14+426)=3492/3932=0.88=88\%$. The Type-I error is the ratio of misclassified cases of actual bankrupt cases declared as non-bankrupt by the model with total bankrupt cases i.e Type-I = $M_1$/$(N_1+M_1$) = $426/(1966+426)=426/2392=0.177=17.7\%$. The Type-II error is the ratio of misclassified cases of actual non-bankrupty cases declared as bankrupt by the model with total non-bankrupty cases i.e Type-II=$M_2$/$(N_2+M_2$)=$14/(14+1526)=14/1540=0.009=0.9\%$.

The proposed method with thresholds as given in Equation \ref{eqn12} is accurate in classifying $88.8\%$ of total samples with Type I error to be only $17\%$ while the Type II error was even better at $0.9\%$. Therefore, there is a positive upward bias which can be addressed by adjusting the thresholds between the credit ratings $A$ and $BBB$ as the boundaries fall in the grey zone. Keeping the other thresholds unchanged, we updated the thresholds of $BBB$ as  $-1.0 < H_{i,j} \leq 0.25$ and $A$ as $0.25 < H_{i,j} \leq 1.5$ from the classification table, we obtained the Type I error as $4\%$ and Type II error as $5\%$ with overall accuracy of $95\%$. To see if the sensitivity be further improved we updated the thresholds of $BBB$ as  $-1.0 < H_{i,j} \leq 0.5$ and $A$ as $0.5< H_{i,j} \leq 1.5$ keeping the others unchanged. We found from the classification table the Type I error as $0.16\%$ whereas Type II error has increased to $21.7\%$ with overall accuracy of $91.4\%$. Therefore, the choice of thresholds for transition from bankruptcy to non-bankruptcy should be chosen with caution so that both Type I and Type II errors are at minimum. 

Even though the samples are disproportionate the Algorithm~\ref{alg1} has out-performed the accuracies obtained using the Altman's Z-score methods.


\section{Conclusions and Discussion}
\label{sec6}
A new nonlinear transformation procedure for building a log linear model for computing the Z-score is proposed. Based on the new Z-score ($Z_M$) a new indexing measure is proposed by fitting the data to a P3 distribution and then obtaining the deviations of the given dataset from the standard normal using an equi-probability transformation. The multivariate discriminate analysis (MDA) for predicting the bankruptcy index has shown that the proposed methodology has given highest accuracy of $98.5\%$ which is higher by $5\%$ as compared with Altman's Z-score. The classification accuracies of the transformed financial  ratios in predicting the bankruptcy is around $93.7\%$ as compared to $87.4\%$ obtained by using the factors of Altman's procedure. The accuracies of the proposed method with P3 distribution was $92.2\%$ where as a model with Pareto distribution resulted in an accuracy of $80\%$. Though the methodology is universal and serves as a generalized tool, there is an immense need to validate with global datasets. Also, mutual interference among financial ratios is an important aspect that requires further investigation. We defer our ongoing work in this direction to a subsequent exposition.

\section*{Acknowledgements}
The second author expresses his gratefulness to Sri B. Sambamurthy, Director, Institute for Development and Research in Banking Technology (IDRBT), Hyderabad, India for his encouragement and support. We express our gratefulness to the anonymous referees for their constructive comments and suggestions. 

\bibliographystyle{aps-nameyear}

\end{document}